\documentclass[11pt]{article}

\let\ssection=\section
\renewcommand{\section}{\setcounter{equation}{0}\ssection}

\def\parag{\hfil\break} %%%%% paragraph
\def\kikezd{\parag\underbar}
\def\p{{\partial}}
\def\vb{{\vec b}}
\def\vx{{\vec x}}
\def\vp{{\vec p}}
\def\vq{{\vec q}}
\def\vj{{\vec{\jmath}}}
\def\vnabla{{\vec\nabla}}

\begin{document}

\setlength{\baselineskip}{16pt}

\title{Galilean symmetry in noncommutative field theory}

\author{
P.~A.~Horv\'athy
\\
Laboratoire de Math\'ematiques et de Physique Th\'eorique\\
Universit\'e de Tours\\
Parc de Grandmont\\
F-37 200 TOURS (France)
\\
L. Martina\\
Dipartimento di Fisica dell'Universit\`a
\\
and\\
Sezione INFN di Lecce. Via Arnesano, CP. 193\\
I-73 100 LECCE (Italy).
\\ and\\
P.~C.~Stichel\\
An der Krebskuhle 21\\
D-33 619 BIELEFELD (Germany)
}

\date{\today}

\maketitle

\begin{abstract}
     When the interaction potential is suitably reordered, the
     Moyal field theory admits two types of  Galilean symmetries,
     namely the conventional mass-parameter-centrally-extended
     one with commuting boosts, but also
     the two-fold centrally extended ``exotic'' Galilean symmetry,
     where the commutator of the boosts yields the noncommutative
     parameter.
     In the free case, one gets an ``exotic''
     two-parameter central extension of the Schr\"odinger group.
     The conformal symmetry is, however, broken by the interaction.
\end{abstract}

%%%%%%%%%%%%%%%%%%%%%%
\section{Introduction}
%%%%%%%%%%%%%%%%%%%%%%

In a recent paper  \cite{Baketal}
Bak et al. consider a scalar field theory on
the noncommutative plane, described by the action
$S=S_{0}+S_{int}^\star=\int\!d^2\vx dt{\cal L}$,
\begin{equation}
     {\cal L}={\cal L}_{0}-V^\star
     =\left(i\bar{\psi}\,\p_{t}\psi+
     \bar{\psi}\frac{\bigtriangleup\psi}{2}\right)
     -\frac{\lambda}{2}\,
     \bar{\psi}\star\bar{\psi}\star\psi\star\psi
     \label{action}
\end{equation}
where the star means the Moyal product associated with
the noncommutative parameter $\theta$.  Although this looks
like a nonrelativistic theory, Bak et al. mention (without proof) that
both the Galilean and scale invariance are lost.
Our aim here is to point out that
the Galilean symmetry can be restored by suitably
reordering the interaction potential.
Then we find that the symmetry
can be implemented in {\it two different ways}:
while the conventional
one yields the usual, one-parameter central extension,
another, ``Moyal-type'' implementation yields the
``exotic'' two-parameter centrally extended
Galilean symmetry, found before in a point particle context \cite{LeLe}.
We confirm that any nontrivial interaction
does indeed break the scale invariance, but
in the free case the symmetry
actually extends to a novel type of ``exotic'' (i. e.
two-parameter-centrally-extended) conformal (Schr\"odinger)
symmetry.
\goodbreak

%%%%%%%%%%%%%%%%%%%%%%%%%%%%%%%%%%%%%%%%%%%%%%
\section{Exotic Galilean symmetry}
%%%%%%%%%%%%%%%%%%%%%%%%%%%%%%%%%%%%%%%%%%%%%%

Let us start with the boosts, whose infinitesimal action on
non-relativistic space-time, $\delta_{b}\vx=\vb t,\,\delta_{b}t=0$,
is conventionally implemented as
\begin{equation}
     \delta^{0}_{b}\psi=i\vb\cdot\vx\,\psi-t\vb\cdot\vnabla\psi.
     \label{convbimp}
\end{equation}
This changes the free part of the Lagrange density in
(\ref{action}) by a surface term,
$\delta{\cal L}_{0}=-t\vb\cdot\vnabla{\cal L}_{0}$.
In the commutative case the general
potential is $V(\rho)$ with $\rho=\bar{\psi}\psi$.
  $\delta^{0}\rho=-t\vb\cdot\vnabla\rho$,
and hence the potential
changes in the same way as the free part,
$
\delta^0V=V'\delta\rho=
-t\vb\cdot\vnabla V.
$
In conclusion,
\begin{equation}
     \delta^0_{b}{\cal L}=-t\vb\cdot\vnabla{\cal L},
     \label{convLagchange}
\end{equation}
implying the Galilean invariance of the action.
In the noncommutative case, however,
the interaction potential in (\ref{action})
is equivalent rather to
\begin{equation}
V^*=(\lambda/2)\,\rho_{r}\rho_{l},
\qquad
\rho_{r}=\bar{\psi}\star\psi,
\qquad
\rho_{l}=\psi\star\bar{\psi}.
\label{ncpot}
\end{equation}
Then the relations
\begin{equation}
     f\star(x_{i}g)
     =
     x_{i}(f\star g)-\frac{i\theta}{2}\,\epsilon_{ij}\,\p_{j}f\star g
     \qquad
     (x_{i}f)\star g
     =
     x_{i}(f\star g)+\frac{i\theta}{2}\,\epsilon_{ij}\,f\star \p_{j}g,
     \label{xrel}
\end{equation}
readily inferred from the definition of the Moyal product allow us
to establish
\begin{equation}
\delta^{0}_{b}\rho_{a}=\pm
\frac{\theta}{2}\vb\times\vnabla\rho_{a}
-t\vb\cdot\vnabla\rho_{a}
\label{ordlrdenschange}
\end{equation}
with the plus sign for $a=r$ and the minus for $a=l$.
Hence the potential changes as
\begin{equation}
     \delta^0_{b}V^*=
     {\theta}\lambda\,\vb\times\Big(
     \vnabla\rho_{r}\star\rho_{l}
     -\rho_{r}\star\vnabla\rho_{l}
     \Big)-t\vb\cdot\vnabla V^*.
     \label{Vstarchange}
\end{equation}
Owing to the sign change above, the first term here is {\it not}
a surface term. The invariance is therefore broken,
as stated  by Bak et al \cite{Baketal}.
\goodbreak

Now we argue that, in the Moyal context,
(\ref{convbimp}) is {\it not} the correct way to act for a boost.
Remember that the imaginary factor in front of $\psi$ is
in fact a[n infinitesimal] ``compensating gauge transformation''
which, in the present context, acts by the Moyal, rather then
by the ordinary multiplication,
$\psi\to g\star\psi$, $g\in U(1)_{*}$, \cite{Szabo}.
(\ref{convbimp}) should therefore be modified as
\begin{equation}
     \begin{array}{ll}
     \delta^{\star}_{b}\psi=i\vb\cdot\vx\star\psi-t\vb\cdot\vnabla\psi
     =i\vb\cdot\vx\,\psi
     -\frac{\theta}{2}\vb\times\vnabla\psi-t\vb\cdot\vnabla\psi
     \\[10pt]
     \delta^{\star}_{b}\bar{\psi}=
     -i\vb\cdot\bar{\psi}\star\vx-t\vb\cdot\vnabla\bar{\psi}
     =-i\vb\cdot\vx\,\bar{\psi}
     -\frac{\theta}{2}\vb\times\vnabla\bar{\psi}
     -t\vb\cdot\vnabla\bar{\psi}.
     \end{array}
     \label{ncbimp}
\end{equation}
The sign change in front of the first term here is consistent
with the formula $\overline{f\star g}=\bar{g}\star\bar{f}$.

Let us first investigate the free theory.
The new implementation (\ref{ncbimp})
changes ${\cal L}_{0}$ again by a surface term,
\begin{equation}
     \delta^{\star}_{b}{\cal L}_{0}=-t\vb\cdot\vnabla{\cal L}_{0}-
\frac{\theta}{2}\vb\times\vnabla{\cal L}_{0},
\label{newfreeactchange}
\end{equation}
cf. (\ref{convLagchange}), so that the free action $S_{0}$ is left invariant.
Hence, the free theory admits our new type
of Galilean symmetry.

The new term in (\ref{ncbimp}) contributes
to the conserved quantity associated through
Noether's theorem, which says that
if ${\cal L}$ changes as
$\delta{\cal L}=\p_{\alpha}K^{\alpha}$ under an
infinitesimal coordinate change $\delta \vx$, then
\begin{equation}
   \int\left(\frac{\delta{\cal L}}{\delta(\p_{t}\psi)}\delta\psi
  +\delta\bar{\psi}\frac{\delta{\cal L}}{\delta(\p_{t}\bar{\psi})}
     -K^{t}\right)d^2\vx
\label{Noether}
\end{equation}
is a constant of the motion.
For a boost, implemented as in (\ref{ncbimp}), we get
\begin{equation}
     {\cal G}_{i}=-\int d^2\vx\,x_{i}\vert\psi\vert^2
     +t{\cal P}_{i}
     +\frac{\theta}{2}\epsilon_{ij}\,{\cal P}_{j}
     \label{exoboost}
\end{equation}
where
${\cal P}_{i}=-i\displaystyle\int d^2x\bar{\psi}\p_{i}\psi$
is the momentum.
Our clue is that the extra piece proportional to $\theta$ changes the
commutator of the boost components,
\begin{equation}
     \big\{{\cal G}_{i},{\cal G}_{j}\big\}=\epsilon_{ij}k,
     \qquad
     k\equiv
     \theta\!\int\! d^2x\vert\psi\vert^2,
     \label{exoticcharge}
\end{equation}
where the Poisson bracket is
\begin{equation}
     \big\{{\cal F}, {\cal G}\big\}=(-i)
     \displaystyle\int d^{2}\vx\left(\frac
     {\delta{\cal F}}{\delta\psi}\frac{\delta{\cal G}}{\delta\bar{\psi}}
     -
     \frac{\delta{\cal G}}{\delta\psi}\frac{\delta{\cal F}}
     {\delta\bar{\psi}}\right).
\label{PBdef}
\end{equation}

Adding the energy, the angular momentum, and the mass,
\begin{eqnarray}
     {\cal H}_{0}=\int d^2\vx\,
     \frac{1}{2}\vert\vnabla\psi\vert^2,
     \\[8pt]
     {\cal J}=-i\int d^2\vx\epsilon_{ij}x_{i}\bar{\psi}\p_{j}\psi,
     \\[8pt]
     M=\int\!d^2\vx\rho_{r}=\int\!d^2\vx|\psi|^2
\label{enerangmommass}
\end{eqnarray}
(also derived by Noether's theorem),
we obtain the  ``exotic''
{\it two-fold centrally extended}
Galilei algebra \cite{LeLe}, whose commutation relations only differ
from those of the usual, singly-extended galilean algebra in that
the boosts yield the second central charge
$k$ in (\ref{exoticcharge}). The usual central term is
the mass $M$, associated with the phase invariance.

It is worth noting that
the conventional implementation (\ref{convbimp})
used by Hagen \cite{Hagen}
yields instead (\ref{exoboost}) without the extra piece,
so that the boost components  commute
\footnote{This corrects an error in calculating
the commutator, committed in \cite{DH}.}.

We conclude that, for a free particle, both (\ref{convbimp}) and
(\ref{ncbimp}) act as symmetries.

Let us mention that
the infinitesimal action associated with a
conserved quantity ${\cal G}$ is
\begin{equation}
     \delta^\star_{{\cal G}}\psi=\Big\{\psi,{\cal G}\Big\}.
     \label{record}
\end{equation}

Interestingly, choosing the equivalent expression
$-i\psi\p_{t}\bar{\psi}$
in the free action, the phase invariance would yield
$\int\!d^2\vx\rho_{l}$ instead of
$\int\!d^2\vx \rho_{r}$; due to the integral property
$\int \!d^2\vx f\star g=\int\!d^2\vx fg$, this is, however,
the same as our previous $M$.
Note also that while the $\rho_{a}$ are
not positive definite, their integral, $M$, is positive.

Both densities satisfy a continuity equation,
\begin{equation}
\p_{t}\rho_{a}+\vnabla\cdot\vj_{a}=0,
\qquad
\vj_{r}=\vnabla\psi\star\bar{\psi}-
\psi\star\vnabla\bar{\psi},
\quad
\vj_{l}=
\bar{\psi}\star\vnabla\psi
-\vnabla\bar{\psi}\star\psi.
\label{current}
\end{equation}
This could be used as the starting point for a
noncommutative hydrodynamics \cite{NChydro}.
Note that while the coordinate-space densities
have complicated commutation relations, in
momentum-space they satisfy the trigonometric
algebra \cite{trigonal}
\begin{equation}
     \Big\{\widetilde{\rho}_{a}(\vq),\widetilde{\rho}_{a}(\vp)\Big\}
     =
    \pm 2\sin\left(\frac{\theta}{2}\vq\times\vp\right)
     \widetilde{\rho}_{a}(\vq+\vp)
     \label{trigonalg}
\end{equation}
with the positive/negative sign for $a=r$ and $a=l$, respectively.
\goodbreak

%%%%%%%%%%%%%%%%%%%%%%%%%%%%%%%%%%
\section{Exotic conformal symmetry}
%%%%%%%%%%%%%%%%%%%%%%%%%%%%%%%%%%

The 2-parameter conformal extension
of the Galilei group  (called the Schr\"odinger group) \cite{JHN}
can now be considered.
The new generators are the dilations and expansions,
implemented infinitesimally according to
\begin{equation}
     \begin{array}{ccc}
    \delta_{\Delta}\vx=\Delta\vx,\quad\hfill
    &\delta_{\Delta}t=2\Delta t,\quad\hfill
    &\delta^0_{\Delta}\psi=
    -\Delta\big[\psi+\vx\cdot\vnabla\psi+2t\p_{t}\psi\big],\hfill
\\[8pt]
\delta_{\kappa}\vx=\kappa t\vx,\,\hfill
&\delta_{\kappa}t=\kappa t^2,\,\hfill
&\delta^0_{\kappa}\psi=-\kappa\big[(-\frac{i}{2}x^2+t)\psi
+t\vx\cdot\vnabla\psi+t^2\p_{t}\psi\big],\hfill
\end{array}
\label{dilexp}
\end{equation}
respectively, where $\Delta>0$ and $\kappa$ is real.
($\Delta\vx$ means $\vx$ multiplied by $\Delta$
and $\Delta t $ means $t$ multiplied by $\Delta$).
More generally, an
infinitesimal Schr\"odinger transformation is of the form
$\delta x_{i}=f_{i}(x,t)$,
$\delta t=g(t)$
with $f_{i}(x,t)=F_{i}(t)+x_{i}G(t)$. When
  implemented conventionally,
\begin{equation}
     \delta^{0}\psi(x,t)=ih(x)\psi(x,t)
     -f_{i}(x,t)\p_{i}\psi(x,t)+\big[k(t)-g(t)\p_{t})\big]\psi(x,t)
     \label{cimplem}
\end{equation}
where the coefficients are suitable real functions,
it leaves invariant ${\cal L}_{0}$ and is therefore a symmetry
for a free field. The associated conserved quantities
span, w. r. t. the Poisson bracket (\ref{PBdef}) the
one-parameter-centrally-extended Schr\"odinger algebra \cite{JHN}.

With hindsight to the noncommutative case, let us
consider instead\footnote{
According to \cite{JP}
  one should take
$-(1/2)\big(f_{i}\star\p_{i}\psi+\p_{i}\psi\star f_{i}\big)$
for the second term. But  when $f_{i}$ is at most linear
in $\vx$, this reduces to our expression.
For $\bar{\psi}$ the first term becomes $-i\bar{\psi}\star h(x)$,
cf. (\ref{ncbimp}).}
  \begin{equation}
     \delta^{\star}\psi(x,t)=ih(x)\star\psi(x,t)
     -f_{i}(x,t)\p_{i}\psi(x,t)+\big[k(t)-g(t)\p_{t}\big]\psi(x,t).
     \label{ncimplem}
\end{equation}
The ``exotic'' dilatations act in the same way
as the conventional ones in (\ref{dilexp});
for the ``exotic'' expansions we find
\begin{equation}
     \delta^*_{\kappa}\psi=-\kappa\big[(-\frac{i}{2}x^2+t)\psi
+t\vx\cdot\vnabla\psi+t^2\p_{t}\psi\big]
-\kappa\left[
\frac{\theta}{2}\vx\times\vnabla\psi+\frac{\theta^2}{4}
\p_{t}\psi\right].
\label{exexp}
\end{equation}

The new transformation is readily seen to be still a symmetry that extends the
``exotic'' Galilei algebra
by adding the two conformal generators \footnote{Note
that (\ref{dilexp}) is also consistent with (\ref{record}).}
\begin{equation}
     \begin{array}{ll}
     {\cal D}=-2t{\cal H}_{0}+\displaystyle\frac{1}{2i}\int d^2\vx x_{i}
     \big(\bar{\psi}\p_{i}\psi-(\p_{i}\bar{\psi})\psi\big)
     \\[8pt]
     {\cal K}=
     t^2{\cal H}_{0}+t{\cal D}-
     \displaystyle\frac{1}{2}\int\!d^2\vx\,\vx^2|\psi|^2
     +\frac{\theta}{2}{\cal J}-\frac{\theta^2}{4}{\cal H}_{0}.
     \label{ncconf}
     \end{array}
\end{equation}

Unlike in the commutative case,
${\cal D}, {\cal H}_{0}$ and ${\cal K}$ do {\it not}
close to an o$(2,1)$ \cite{JHN} but yield
\begin{equation}
	\{{\cal D},{\cal H}_{0}\}=2{\cal H}_{0}
	\qquad
	\{{\cal D},{\cal K}\}=-2{\cal K}+\theta{\cal J}-\theta^2{\cal H}_{0},
	\qquad
	\{{\cal H}_{0},{\cal K}\}={\cal D}.
\label{ncconfrel}
\end{equation}
When added to the Galilean generators we get a closed
algebra, though, since the non-vanishing brackets read
\begin{equation}
     \{{\cal K},{\cal G}_{i}\}=\theta\epsilon_{ij}{\cal G}_{i},
     \qquad
     \{{\cal K},{\cal P}_{i}\}={\cal G}_{i},
     \qquad
     \{{\cal D},{\cal G}_{i}\}=-{\cal G}_{i}+\theta\epsilon_{ij}{\cal P}_{j}.
     \label{mixed}
\end{equation}

The relations (\ref{exoticcharge}-\ref{ncconfrel}-\ref{mixed})
define a new, two-parameter central extension of the Schr\"odinger
algebra which seems to have escaped attention so far.
We call it the ``exotic Schr\"odinger algebra''.

%%%%%%%%%%%%%%%%%%%%%%%%%%%%%%%%%%%%%%%%%%%%%%
\section{Symmetry properties of the potential}
%%%%%%%%%%%%%%%%%%%%%%%%%%%%%%%%%%%%%%%%%%%%%%

We now turn our attention to the potential.
  Let us first consider a boost, implemented ``exotically''
i. e. as in (\ref{ncbimp}).
It follows readily [cf. (\ref{ordlrdenschange})]
that the right and left densities  transform differently,
\begin{equation}
     \begin{array}{ll}
\delta^{\star}_{b}\rho_{r}=-t\vb\cdot\vnabla\rho_{r},
\\[8pt]
\delta^{\star}_{b}\rho_{l}=
-t\vb\cdot\vnabla\rho_{l}
-\theta\vb\times\vnabla\rho_{l}.
\end{array}
\label{denschange}
\end{equation}
Hence, for $V^*=\rho_{r}\rho_{l}$ we get
\begin{equation}
\delta^{\star}_{b}V^{\star}
=-t\vb\cdot\vnabla V^*_{r}
-\theta\vb\times\big(\rho_{r}\star\vnabla\rho_{l}\big)
\end{equation}
  Here the term proportional to $\theta$ is
not exact. Therefore, the Galilean invariance
is broken by the potential $V^{\star}$
even for the new implementation
(\ref{ncbimp}).

Let us remember, however, how the potential comes about \cite{LiPi}.
One starts with the second-quantized  expression for the interaction,
\begin{equation}
     \int
     d^2\vx d^2\vx'\bar{\psi}(\vx)\bar{\psi}(\vx')U(\vx-\vx')
     \psi(\vx)\psi(\vx')
     \label{2ndquantint}
\end{equation}
where $U$ is a two-body potential. Choosing the contact interaction
$U=(\lambda/2)\delta(\vx-\vx')$ yields a quartic $V=(\lambda/2)\rho^2$.

Promoting the commutative theory into a noncommuting one
requires to replacing the ordinary products by Moyal products.
This requires particular care, though. For example, putting
naively Moyal stars between the various factors in
(\ref{2ndquantint}) while keeping the original order
would lead to inconsistency: expressions of the
form $\psi(\vx)\star\psi(\vx')$ that should be defined
due to associativity, would require us to
redefine the Moyal product.
Our clue is that this procedure is {\it ambiguous},
as the order of the factors is irrelevant in the
commutative theory, but not in its Moyal version.
  Rearranging as, e. g.,
\begin{equation}
     \int
     d^2\vx d^2\vx'
     \bar{\psi}(\vx)\star\psi(\vx)
     \star U(\vx-\vx')\star\bar{\psi}(\vx')\star\psi(\vx')
     \label{rearrangedpot}
\end{equation}
(where the various products {\it are} well-defined)
would yield, instead of $V^*$ in (\ref{action}),
\begin{equation}
     \widetilde{V}^*=\frac{\lambda}{2}
     \bar{\psi}\star\psi\star\bar{\psi}\star\psi
     \label{ourpot}
\end{equation}
equivalent to $(\lambda/2)\rho_{r}^2$ or to
$\widetilde{\widetilde{V}}^*=(\lambda/2)\rho_{l}^2$
[which could also be obtained by a suitable reordering].
Remarkably, it is this expression that had been used by
Lozano et al. in their noncommutative nonrelativistic
Chern-Simons vortex construction \cite{Schap},
and also by Langmann et al. \cite{Langmann}
in their recent exact scalar field solution in a
background magnetic field.

The important fact for us is that the new interaction
{\it is} Galilei invariant, since, by
(\ref{denschange}),
\begin{equation}
     \delta^*_{b}\widetilde{V}^*=-t\vb\cdot\vnabla\widetilde{V}^*.
\end{equation}
Similarly, for $\widetilde{\widetilde{V}}^*$ we get
$\delta^*_{b}\widetilde{\widetilde{V}}^*=-t\vb\cdot\vnabla
\widetilde{\widetilde{V}}
-\theta\vb\times\vnabla\widetilde{\widetilde{V}}^*$
which is again a surface term, so that the Galilean symmetry is
again restored.
Clearly, the same statement holds for any ``pure'' function
of $\rho_{r}$ or of $\rho_{l}$ alone.  For a mixture,
$\rho=\epsilon_{r}\rho_{r}+\epsilon_{l}\rho_{l}$
where the $\epsilon_{i}$ are real coefficients, the cross
term would break, however the symmetry,
whenever $\epsilon_{r}\epsilon_{l}\neq0$.

We failed to find a natural way to reproduce the potential
$V^*=\bar{\psi}\star\bar{\psi}\star{\psi}\star\psi$
in (\ref{action}),
proposed by Bak et al.  \cite{Baketal}\footnote{Their potential
could be obtained by inserting further $\delta$ factors.}.
Another argument in favor of our choice (\ref{ourpot})
is that it is, unlike that of Bak et al., $V^*$, renormalizable
as well as invariant w. r. t. deformed
$U(1)$ transformations \cite{Arefeva}, p. 25.
It is, therefore, (\ref{ourpot}) that we shall adopt in what follows.

Interestingly, the modified potential $\widetilde{V}^*$ also
allows the conventional symmetry (\ref{convbimp})\footnote{
But this is not a surprise as (\ref{convbimp}) and (\ref{ncbimp}) just
differ by a space-translation term.}.
(\ref{ordlrdenschange})  indeed implies that
\begin{equation}
     \delta^0_{b}\rho_{a}^2=
     \pm\frac{\theta}{2}\vb\times\vnabla\rho_{a}^2
     -t\vb\cdot\vnabla\rho_{a}^2
\end{equation}
$a=r,l$. Finally, the {\it conventional} interaction
in terms of  $\rho$ is also $\delta^*$-invariant,
proving the Galilean symmetry also in this case.
In conclusion, any of the ``pure'' expressions
$
V^*(\rho_{a})
$
as well as the standard potential $V=\rho^2$
provide us with a theory which is {\it Galilei-invariant in two ways}~:
the conventional implementation yields the usual one-parameter
extension with commuting boosts, and
the ``star-implementation'' yields the  exotic two-parameter
extension with noncommuting boosts.

Let us the record for completeness the equation of motion
associated with our potential (\ref{ourpot})~: either by variation or
using the Hamiltonian structure, we get the ``Moyalized'' non-linear
Schr\"odinger equation
\begin{equation}
i\p_{t}\psi=-\frac{\bigtriangleup}{2}\psi
+\lambda\,\rho_{l}\star\psi=
-\frac{\bigtriangleup}{2}\psi
+\lambda\,\psi\star\rho_{r}.
\label{NLS}
\end{equation}
For comparision, for the choice $V^*$ of Bak et al.,
the nonlinear term is
$(\lambda/2)(\rho_{r}\star\psi+\psi\star\rho_{l})$.
The commutative counterpart of (\ref{NLS}) is known to be
non-integrable; coupling our system to an external magnetic field,
yields exact solutions \cite{Langmann}, however.

%%%%%%%%%% %%%%%% %%%%%% %%%%%% %%%%%%
%\subsection{Conformal symmetry with potential}
%%%%%%%%%% %%%%%% %%%%%% %%%%%% %%%%%%

The ``ordinary'' conformal symmetry is also consistent with adding a
quartic potential $V(\psi)=\rho^2$.
In fact, using (\ref{dilexp}) we find that
$\delta^0_{\Delta}\rho=\Delta\big(2\rho
-\vx\cdot\vnabla\rho-2t\p_{t}\rho\big)$
and
$\delta^0_{\kappa}\rho=\kappa\big(2\rho
-\vx\cdot\vnabla(t\rho)-t^2\p_{t}\rho\big)$,
so that
\begin{equation}
     \begin{array}{ccc}
	\delta^0_{\Delta}\rho^{2}=
&-\vnabla\cdot(\Delta\vx\rho^2)-\p_{t}(\Delta 2t\rho^2)\quad\hfill
&\hbox{for a dilatation}\ \hfill
\\[10pt]
\delta^0_{\kappa}\rho^{2}=
&-\vnabla\cdot(\kappa t\vx\rho^2)-\p_{t}(\kappa t^2\rho^2)\hfill
&\hbox{for an expansion}\ \hfill
\end{array}\quad.
\label{confondens}
\end{equation}

The associated conserved quantities, that still
form a representation of the one-parameter centrally
extended Schr\"odinger algebra,
only differ from the free expressions in that
${\cal H}_{0}$ is replaced by ${\cal H}={\cal H}_{0}+V$ \cite{JBeg}.

As the ``exotic'' action of a dilatation
is the same as the conventional one
in (\ref{dilexp}) we find, using (\ref{xrel}),
\begin{equation}
     \delta_{\Delta}^*\rho_{r}=\delta_{\Delta}^0\rho_{r}
     =\Delta\left\{
     2\rho_{r}-\vx\cdot\vnabla\rho_{r}-2t\p_{t}\rho_{r}\right\}
     -\Delta i\theta\epsilon_{ij}\p_{i}\bar{\psi}\star\p_{j}\psi
     \label{exodilondens}
\end{equation}
which differs from $\delta^*_{\Delta}\rho$ in the second term
behind $\theta$. Thus, owing precisely to this term,
$\rho_{r}^2$ can not change by a surface term.
The scaling symmetry is
therefore broken,  as stated in \cite{Baketal}.
(The same statement holds also for $\rho_{l}^2$
and $\rho_{r}\rho_{l}$.)
More generally, (\ref{ncimplem})  readily implies that
\begin{equation}
     \begin{array}{ll}
\delta^{\star}\rho_{r}
=
(2k-g\p_{t})\rho_{r}-f_{i}\p_{i}\rho_{r}
-\Big[(f_{i}\p_{i}\bar{\psi})\star\psi
+\bar{\psi}\star (f_{i}\p_{i}\psi)\Big]
\\[8pt]
\delta^{\star}\rho_{l}=
(2k-g\p_{t})\rho_{l}-f_{i}\p_{i}\rho_{l}
-\Big[(f_{i}\p_{i}\psi)\star\bar{\psi}
+\psi\star (f_{i}\p_{i}\bar{\psi})\Big].
\end{array}
\end{equation}
As the second brackets involve here expressions of the form
\begin{eqnarray*}
     x_{i}\p_{i}\rho\pm i\theta\epsilon_{ij}
\p_{i}\bar{\psi}\p_{j}{\psi}+O(\theta^2)
\end{eqnarray*}
the additional terms are not given by a surface term.
that never vanish unless $G(t)=0$.
In conclusion, any potential made of products of
$\rho_{r}$ and $\rho_{l}$  necessarily
breaks the conformal invariance.

%%%%%%%%%%%%%%%%%%%%
\section{Discussion}
%%%%%%%%%%%%%%%%%%%%

An interesting feature of the model studied here
is the two--way Galilean symmetry, and one can be
puzzled how this can happen. Let us consider the ``Moyalized''
counterpart of ${\cal L}_{0}$, namely
\begin{equation}
     {\cal L}_{0}^\star
     =i\bar{\psi}\star\p_{t}\psi-
     \frac{1}{2}\vnabla\psi\star\vnabla\bar{\psi}
     \label{staraction}
\end{equation}
Now the conventional implementation (\ref{convbimp})
of the boosts is natural for ${\cal L}_{0}$,
as is (\ref{ncbimp})
for ${\cal L}_{0}^\star$. But the integral property
$\int\! f\star g=\int\! fg$ implies  that ${\cal L}_{0}$
and ${\cal L}_{0}^\star$ are equivalent, so one can use either
of them to describe a free field.
This resolves the paradox which says that
noncommutativity does not alter the free theory.

\kikezd{Acknowledgement}.
We are indebted to F. Schaposnik
for enlightening discussions and to R. J. Szabo
for correspondence. One of us (P. H.)
would like to thank for hospitality
the University of Lecce.

%%%%%%%%%%%%%%%%%%%%%%%%%%%%%%%%%%%%%%%%%%%%%%%%%%%%%%%%%%%%%%%%%%%%%%%%%%%%%%
%%%%%%%%%%%%%%%%%%%%%%%%%%%%%%%%%%%%%%%%%%%%%%%%%%%%%%%%%%%%%%%%%%%%%%%%%%%%%%


\begin{thebibliography}{99}
%%%%%%%%%%%%%%%%%%%%%%%%%%%%%%%%%%%%%%%%%%%%%%%%%%%%%%%%%%%%%%%%%%%%%%%%%%%%%%
%%%%%%%%%%%%%%%%%%%%%%%%%%%%%%%%%%%%%%%%%%%%%%%%%%%%%%%%%%%%%%%%%%%%%%%%%%%%%%
\bibitem{Baketal}
D.~Bak, S.~K.~Kim, K.-S. Soh, and J. H. Yee,
{\sl Phys. Rev. Lett}. {\bf 85}, 3087 (2000).
Analogous expressions were considered before, e. g., by
J.~Gomis, K. Landsteiner, and E. Lopez,
{\sl Phys. Rev.} {\bf D62}, 105006 (2000);
G.~S. Lozano, E. F. Moreno, F. A. Schaposnik,
{\sl Journ. High Energy Phys}. {\bf 02} (2001) 036,
V.~P.~Nair and A.~Polychronakos,
{\sl Phys. Lett}. {\bf B505}, 267 (2001).

\bibitem{LeLe}
J.-M.~L\'evy-Leblond,
{\it Galilei group and Galilean invariance}.
in {\it Group Theory and Applications} (Loebl Ed.),
{\bf II}, Acad. Press, New York, p. 222 (1972);
A. Ballesteros, N.~Gadella and M.~del Olmo,
{\sl Journ. Math. Phys.} {\bf 33}, 3379 (1992);
Y.~Brihaye, C.~Gonera, S.~Giller and P.~Kosi\'nski,
\texttt{hep-th/9503046} (unpublished);
D.~R.~Grigore, {\sl Journ. Math. Phys.} {\bf 37}, 240 and
   {\sl ibid}. {\bf 37} 460 (1996).
For new developments, see
J.~Lukierski, P.~C.~Stichel, W.~J.~Zakrzewski,
   {\sl Annals of Physics (N. Y.)} {\bf 260}, 224 (1997), and
   \texttt{hep-th/0207149}, {\sl Annals of Physics (N. Y.)}
   (in press).
C.~ Duval and P.~A.~Horv\'athy,
{\sl Phys. Lett.} {\bf B 479}, 284 (2000). %[\texttt{hep-th/0002233}];

\bibitem{Szabo}
R.~J. Szabo,
{\it Quantum Field Theory on noncommutative spaces}.
{\sl Phys. Rep.} {\bf 378}, 203 (2003)
[\texttt{hep-th/0109162}].

\bibitem{Hagen}
C.~R. Hagen, {\sl Phys. Lett.} {\bf B539}, 168 (2002).

\bibitem{DH}
C.~ Duval and P.~A.~Horv\'athy,
{\sl Phys. Lett.} {\bf B547}, 306 (2002).

\bibitem{NChydro}
C.~Duval, Z.~Horv\'ath, P.~A.~Horv\'athy,
  {\sl Int. Journ. Mod. Phys.} {\bf  B15}, 3397 (2001);
  %[\texttt{cond-mat/0101449}].
Z.~Guralnik, R.~Jackiw, S.~Y.~Pi,
and A.~P.~Polychronakos,
{\sl Phys. Lett. } {\bf B517}, 450 (2001).
  %[\texttt{hep-th/0106044}].



\bibitem{trigonal}
D.~B. Fairlie, P.~Fletcher and C. K. Zachos,
{\sl Phys. Lett.} {\bf B218}, 203 (1989);
D.~B. Fairlie and C. K. Zachos,
{\sl Phys. Lett.} {\bf B224}, 101 (1989);
J.~Martinez and M.~Stone,
{\sl Int. Journ. Mod. Phys.} {\bf B26}, 4389 (1993).

\bibitem{JHN}
R.~Jackiw, {\sl Physics Today} {\bf 25}, 23 (1980);
U.~Niederer, {\sl Helvetica Physica Acta} {\bf 45}, 802 (1972);
C.~R.~Hagen, {\sl Phys. Rev}. {\bf D5}, 377 (1972).

\bibitem{JP}
R.~Jackiw and S.-Y. Pi,
{\sl Phys. Rev. Lett}. {\bf 88}, 111603 (2002).

\bibitem{JBeg}
R.~Jackiw,
{\it Delta function potential in two and three dimensional
quantum mechanics}. In {\it B\'eg Memorial Volume},
A. Ali and P. Hoodboy eds. (World Scientific, Singapore, 1991).

\bibitem{LiPi}
E.~M.~Lifshitz an L. P.~Pitaevskii,
{\it Statistical Physics}. Part 2. %Theory of the condensed state.
  Landau-Lifshitz Course of Theoretical Physics,
Vol. 9. Chapter II. Sec. 7. Pergamon Press: Oxford (1980).

\bibitem{Schap}
See Lozano, Moreno and Schaposnik in
Ref. \cite{Baketal}.

\bibitem{Langmann}
  E.~Langmann, R.~J.~Szabo,
{\sl Phys. Lett.} {\bf B533}, 168 (2002);
  E.~Langmann, {\sl Nucl. Phys}. {\bf B654}, 404 (2003);
E.~Langmann, R.~J.~Szabo, and K. Zarembo,
\texttt{hep-th/0308082}.

\bibitem{Arefeva}
I.Ya.~Aref'eva, D. M. Belov, A. A. Giryavets,
A.S. Khoshelev and P. B. Medvedev,
{\it Noncommutative field theories and (super)string
field theories}. \texttt{hep-th/0111208}.

\end{thebibliography}
\end{document}